\documentclass[aps,twocolumn,showpacs,preprintnumbers,nofootinbib,prd,10pt,superscriptaddress]{revtex4-2}

\DeclareRobustCommand{\ion}[2]{\textup{#1\,\textsc{\lowercase{#2}}}}

\usepackage{booktabs}
\usepackage{graphicx}
\usepackage{dcolumn}
\usepackage{bm}

\usepackage{amsmath}
\numberwithin{equation}{section}
\usepackage{ragged2e}

\usepackage{amsfonts,amssymb,mathtools,siunitx}
\usepackage[colorlinks=true,     linkcolor=blue, citecolor=blue, filecolor=blue, urlcolor=blue]{hyperref}

\usepackage{natbib}
\usepackage{txfonts}
\usepackage{multirow}
\usepackage{float}
\usepackage{orcidlink}
\usepackage{comment}
\usepackage{tablefootnote}
\usepackage[flushleft]{threeparttable}

\begin{document}


\title{Novel method to trace the dark matter density profile around supermassive black holes with AGN reverberation mapping}

\author{Mayank Sharma\textsuperscript{\ddag}}
\email{mayanksh@vt.edu}
\affiliation{Department of Physics, Virginia Tech, Blacksburg, VA 24061, USA}
\author{Gonzalo Herrera\textsuperscript{\ddag}}
\email{gonzaloh@mit.edu}
\affiliation{Department of Physics and Kavli Institute for Astrophysics and Space Research, Massachusetts Institute of Technology, Cambridge, MA 02139, USA}
\affiliation{Harvard University, Department of Physics and Laboratory for Particle Physics and Cosmology, Cambridge, MA 02138, USA}
\affiliation{Center for Neutrino Physics, Department of Physics, Virginia Tech, Blacksburg, VA 24061, USA}
\author{Nahum Arav}
\affiliation{Department of Physics, Virginia Tech, Blacksburg, VA 24061, USA}
\author{Shunsaku Horiuchi}
\affiliation{Department of Physics, Institute of Science Tokyo, 2-12-1 Ookayama, Meguro-ku, Tokyo 152-8551, Japan}
\affiliation{Center for Neutrino Physics, Department of Physics, Virginia Tech, Blacksburg, VA 24061, USA}
\affiliation{Kavli IPMU (WPI), UTIAS, The University of Tokyo, Kashiwa, Chiba 277-8583, Japan}

\thanks{These authors contributed equally to this work.}

\date{\today}

\begin{abstract}
We propose a new method to determine the dark matter density profile in the vicinity of distant supermassive black holes (SMBH) using reverberation mapping (RM) measurements of active galactic nuclei (AGN). The mapping of multiple emission lines allows the measurement of the enclosed mass within different radii from the central SMBH, which can be used to infer or constrain the dark matter density profile on sub-parsec scales. We apply a toy model based on this method to a sample of fourteen AGN to test its feasibility based on current measurements. We find that for five objects, the observed enclosed mass does grow with radii, hinting towards the presence of a dark matter component at the 1-2 $\sigma$ level. For these sources, we find global evidence for a universal dark matter profile with a preferred radial steepness of index $\gamma \sim 1.6$, consistent with the scenario expected for a dark matter spike mildly relaxed by stellar heating processes. The enclosed dark matter mass, however, is found to be significantly larger than expected. We show that the current RM based mass measurements suffer from large systematic uncertainties, that limit the effectiveness of our method. Our work emphasizes the importance of applying the recent developments in mass determination techniques to target multiple emission lines with future RM and interferometry campaigns. This provides the most direct way of constraining the dark matter density in the sub-parsec regions around extragalactic SMBHs, which is crucial to our understanding of the dynamics and nature of dark matter.

\end{abstract}

\maketitle


\section{Introduction}
Almost all large galaxies (including our Milky Way) are now believed to harbor a Supermassive Black Hole (SMBH) at their center. Constraining the mass of the SMBH is essential in understanding the physical processes that (a) fuel it and lead to its evolution and (b) affect both its immediate and large-scale surroundings. Our current best measurements for the mass of a SMBH come from within our own galaxy, where precise tracking of the orbits of individual stars around Sagittarius A* (with effective spatial resolution of 6 $\times$ $10^{-4}$ pc) constrain its mass as $4.4^{+0.4}_{-0.4}$ $\times$ $10^6M_{\odot}$ \citep{genzel2010galactic}. Even for the next nearest set of galaxies, their large distance from us limits our ability to resolve individual stellar orbits. Thus our estimates for BH masses in these galaxies come from dynamical modeling of the bulk motions of stars and gases in their sphere of influence \citep[see][for a review of prominent $M_{\rm BH}$ determination techniques]{kormendy2013coevolution}.\par
Besides harboring a SMBH at their center, it is also believed that galaxies should contain a dark matter halo extending to kpc scales from their center \cite{Sofue:2000jx,Read:2014qva,Bullock:2017xww,salucci2019distribution}.
One pressing problem in astrophysics and cosmology resides in the yet unknown fundamental nature of the dark matter composing these halos. Early simulations of $\Lambda$ cold dark matter (CDM) suggest that the dark matter distribution in galaxies may follow a universal form, growing roughly as $\propto r^{-1}$ towards the center of the galaxies \citep{Navarro:1995iw, Navarro:1996gj}. Later simulations including baryonic feedback suggest instead that the dark matter density profile may be shallower towards the inner regions of the galaxy, aligning with the observations from rotational curves for some galaxies \citep{Moore:1994yx, de_Blok_2009}. Current discrepancies only apply down to roughly $\sim 1$ kpc as for smaller radii numerical simulations are less reliable and observational methods are uncertain, leaving the nature of dark matter at smaller scales largely unknown. Inner still, in the close vicinity of the SMBH, the dark matter profile is expected to be modified by the strong gravitational potential of the black hole. Models of adiabatically accreting black holes, in particular, predict a significant enhancement of the dark matter density in the region of gravitational influence, forming a so called ``dark matter spike" \citep{Peebles,BahcallWolff,Young1980,Quinlan:1994ed,Gondolo:1999ef}. Adiabatic growth is indeed expected to occur for most SMBHs with masses below $M_{\rm BH}\lesssim 10^{10}$M$_{\odot}$ \citep{Sigurdsson:2003wu, Herrera:2023nww}. If present, such structures could have profound implications on our understanding of the fundamental nature of dark matter. For instance, the presence of a density spike could facilitate the annihilation of dark matter particles into observable gamma-rays \citep{Gondolo:1999ef}. Similarly the spike would also enhance the scatterings of dark matter with cosmic rays, gamma-rays and neutrinos in these environments, altering their observable astrophysical fluxes \citep{Gorchtein:2010xa,Herrera:2023nww, Ferrer:2022kei, Cline:2022qld}. In addition, the spike may also modify gravitational wave signatures from binary mergers \citep{Kavanagh:2020cfn,Bertone:2024rxe, Alonso-Alvarez:2024gdz}. \par
So far, there exist robust constraints only on the dark matter spike around Saggitarius A$^{*}$, from the dynamics of the star S2, whose orbit passes close to the central black hole \citep{Lacroix:2018zmg, Shen:2023kkm}\footnote{Recently, the possibility of constraining the dark matter spike around the SMBH binary OJ 287 from its orbital period has been discussed \citep{chan2024robustevidenceshowingdark, Deb:2025raq}.}. These constraints are mild, only able to test optimistic dark matter spike scenarios. 
For active galaxies at cosmic distances, however, no constraints on the dark matter spike around their central SMBH exist, since there is not enough resolution to resolve stellar or gaseous orbits within sub-parsec distances from the central black hole. We propose a novel method which aims to remedy this, inferring the dark matter distribution in the vicinity of distant SMBHs based on reverberation mapping (RM) measurements. \par
RM is a technique that relies on spectroscopic monitoring of active galactic nuclei (AGN). It maps the response of prominent broad emission line features in AGN spectra to changes in their underlying continuum emission, arising from the accretion disk. The delay between the two responses is then an indicator of the location of the gas in the so-called broad line region (BLR). Importantly, the derived scale length for different emission lines vary, with lines that are characteristic of high-ionization gas occurring closer
to the central black holes than those characteristic of low-ionization gas \citep{Peterson:2004am}. The distance of the gas from the continuum source ($r_c$), combined with the velocity dispersion of the emission line ($\Delta V$), is used to infer the mass of the central black hole as:  
\begin{equation}\label{eq:rev_map1}
M_{\mathrm{BH}}=\frac{f r_{c} \Delta V^2}{G}
\end{equation}
where $G$ is the gravitational constant and $f$ is a fudge factor of order unity that depends on the largely unknown geometry and kinematics of the BLR, and is also known as the virial coefficient. While it is possible to obtain the value of $f$ for individual AGN, it requires detailed dynamical modeling of the BLR using high-quality, velocity-resolved measurements and has only been achieved in the case of a few objects \citep[e.g.,][]{pancoast2014modelling, williams2018lick, Villafa2022}. In most cases, therefore, an average value of $f$ is used, which is typically obtained by using the relationship between the black hole mass and stellar velocity (i.e., $M_{BH}-\sigma_{\star}$ relationship) in local galaxies \citep[e.g.,][]{onken2004supermassive, graham2011expanded,grier2013stellar}. These studies have revealed a range of values for $f$ between $\sim$ 2.8 and 5.5, indicating a typical systematic uncertainty of a factor of 2 in individual RM-based black hole masses.\par 
We make the important observation here that equation (\ref{eq:rev_map1}) is actually an estimate of not just $M_{BH}$, but the \textit {total enclosed mass} within the derived distance scale from each line, such that    
\begin{equation}\label{eq:rev_map}
M_{\rm BH}+\int_{R_S}^{r_c} 4\pi r^2(\rho_{\rm DM}(r)+\rho_b(r)) dr=\frac{f r_c \Delta V^2}{G},
\end{equation}
where $\rho_{\rm DM}$ and $\rho_{\rm b}$ are the dark matter and baryonic density profiles in the region, respectively, and $R_S$ is the Schwarzschild radius of the black hole. The total baryonic mass in the region of interest consists of the mass of the accretion disk and the BLR and is expected to be of the order of  $\sim$ 1 percent of the mass of the SMBH \citep{goodman2003self, baldwin2003mass}. The contribution of $\rho_{\rm b}$ to the total enclosed mass can thus be ignored. Using equation (\ref{eq:rev_map}) for a sample of AGN where the enclosed mass has been obtained from RM measurements of multiple emission lines, we are able to investigate for the first time, the radial dependence of the dark matter density profile in the vicinity of distant SMBHs. \par
The paper is organized as follows. In Section \ref{sec:DM_profile}, we discuss theoretical expectations on the dark matter density profile around SMBHs. In Section \ref{sec:sample}, we discuss our sample selection of AGN and the methodology used. In Section \ref{sec:results}, we present our results. Specifically, we first confront our sample data with a flat constant black hole mass versus an increasing mass profile with two free parameters (normalization and slope), and determine which case fits better each object individually. Then, for those objects for which there is preferred evidence for an increasing mass, we find the 1$\sigma$ and 2$\sigma$ contours on the dark matter profile normalization and steepness. We also perform a global analysis of all sources, marginalizing over the normalization of the dark matter density profile, and find the preferred steepness in our sample. In section \ref{sec:discussion}, we discuss the physical plausibility of our results and how they are impacted by the systematic uncertainties in RM measurements. Our conclusions are presented in section \ref{sec:conclusions}.

\section{Dark matter density profile around black holes}\label{sec:DM_profile}
Adiabatically growing black holes are expected to enhance the surrounding dark matter density profile in their region of gravitational influence \citep{Quinlan:1994ed, Gondolo:1999ef}. It has been shown that an initial density profile with radial steepness $\propto r^{-\gamma}$ evolves into a spike with steepness $ \gamma_{\rm sp}=(9-2\gamma)/(4-\gamma)$. For dark matter profiles of galaxies, N-body simulations and kinematical tracers predict the range $\gamma \simeq 0.1-2$, hence the corresponding dark matter spike would vary in the range $\gamma_{\rm sp}=2.25-2.5$. However, it has been argued that some astrophysical processes could redistribute the dark matter spike after its formation, leading to shallower profiles. One possible process is given by hierarchical galaxy mergers \citep{Merritt:2002vj, Milosavljevic:2001vi}, which could reduce the spike steepness to $\gamma_{\rm sp} \sim 0.5$. However, it was shown in \cite{Merritt:2006mt} and \cite{Shapiro:2022prq} that the dark matter spikes could collisionally regenerate via gravitational scattering with stars \citep[referred to as stellar heating;][]{Merritt:2003qk, Gnedin:2003rj}, leading to $\gamma_{\rm sp} >1.5$. It is also possible that the black hole is displaced from the gravitational center of the dark matter halo, which may also lead to shallower indices than a spike \citep{Ullio:2001fb}. Given the variety of possible profiles suggested by semi-analytical models, here we parametrize the dark matter density profile around the SMBH simply as
\begin{equation}\label{eq:parametrization}
\rho_{\rm DM}(r)=\rho_0\left(\frac{r}{0.1 \textrm{ pc}}\right)^{-\gamma} \, \,\, \mathrm{for} \, \,\, 2R_S<r<R_{\rm sp}.
\end{equation}
which depends on the normalization factor $\rho_0$ and the steepness $\gamma$. For $r < 2R_s$ the dark matter density vanishes \citep[as shown by][]{Sadeghian:2013laa}, whereas for  $r > R_{\mathrm{sp}}$, defined as the extension of the spike, the density profile is restored to its initial form. We restrict our analysis to the empirical data obtained from RM measurements, such that we can only derive constraints on the extension of the spike ($R_{\rm sp}$) up to the farthest data point in each source. In reality, theoretical models predict $R_{\rm sp} \sim 10^{-2}-1$ pc, but such predictions are correlated with $\rho_0$ and $\gamma$ \citep{Gondolo:1999ef}, which are independent parameters in our analysis.
The normalization of the dark matter density profile, quantified by $\rho_0$, is expected to be correlated with $M_{\rm BH}$, thus, a global inference of this value from all the sources we considered is not useful. We thus leave $\rho_0$ as a separate free parameter for each source. The steepness of the dark matter profile, $\gamma$, may be treated as an universal variable for all sources, since it does not depend on the mass of SMBH in adiabatic formation scenarios. However, since the merger history and activity of the galaxy may impact its value, we do not enforce the same value for all the sources considered.

\begingroup
\setlength{\tabcolsep}{5pt}
\renewcommand{\arraystretch}{1.09}
\begin{table}
\caption{Compiled multi-line mass determinations.\label{table:emdata}}
\centering
\begin{tabular}{ccccc}
\hline
Source & Lines & $r_c$ & $M_{\rm BH}$  & Ref \\
&  & (lt. days) & ($10^7$ $M_{\odot}$) &  \\
\hline
\multirow{2}{*}{Mrk 335}& \ion{He}{II} & $2.7^{+0.6}_{-0.6}$ &$2.6^{+0.8}_{-0.8}$ & \multirow{2}{*}{ \cite{grier2011reverberation}} \\
& \ion{H}{$\beta$}  & $13.9^{+0.9}_{-0.9}$ &$2.7^{+0.3}_{-0.3}$ &  \\
\hline
\multirow{5}{*}{3C 120}& \ion{He}{II} & $12.0^{+7.5}_{-7.0}$ &$17.3^{+10.5}_{-10.5}$ & \multirow{5}{*}{\cite{kollatschny2014broad}} \\
& \ion{H}{$\gamma$}  & $23.9^{+4.6}_{-3.9}$ &$19.4^{+8.9}_{-8.9}$ \\
& \ion{He}{I}  & $26.8^{+6.7}_{-7.3}$ &$14.8^{+4.6}_{-4.6}$ \\
& \ion{H}{$\beta$}  & $27.9^{+7.1}_{-5.9}$ &$10.8^{+2.6}_{-2.6}$ \\
& \ion{H}{$\alpha$}  & $28.5^{+9.0}_{-8.5}$ &$7.7^{+2.3}_{-2.3}$ \\
\hline
\multirow{3}{*}{MCG+08-11-011}& \ion{He}{II} & $1.21^{+0.29}_{-0.33}$ &$0.63^{+1.28}_{-0.42}$ & \multirow{3}{*}{\cite{fausnaugh2017reverberation}} \\
& \ion{H}{$\gamma$}  & $12.38^{+0.46}_{-0.49}$ &$2.76^{+5.37}_{-1.82}$ \\
& \ion{H}{$\beta$}  & $14.98^{+0.34}_{-0.28}$ &$2.82^{+5.50}_{-1.86}$ \\
\hline
\multirow{3}{*}{NGC 2617}& \ion{He}{II} & $1.75^{+0.34}_{-0.38}$ &$0.62^{+1.43}_{-0.43}$ & \multirow{3}{*}{\cite{fausnaugh2017reverberation}} \\
& \ion{H}{$\gamma$}  & $0.81^{+0.59}_{-0.61}$ &$0.66^{+2.16}_{-0.51}$ \\
& \ion{H}{$\beta$}  & $6.38^{+0.44}_{-0.50}$ &$3.24^{+6.32}_{-2.14}$ \\
\hline
\multirow{4}{*}{Mrk 110}& \ion{He}{II} & $3.9^{+2.8}_{-0.7}$ &$2.25^{+1.63}_{-0.45}$ & \multirow{4}{*}{\cite{kollatschny2001short}} \\
& \ion{He}{I}  & $10.7^{+8.0}_{-6.0}$ &$1.81^{+1.36}_{-1.03}$ \\
& \ion{H}{$\beta$}  & $24.2^{+3.7}_{-3.3}$ &$1.63^{+0.33}_{-0.31}$ \\
& \ion{H}{$\alpha$}  & $32.3^{+4.3}_{-4.9}$ &$1.64^{+0.33}_{-0.35}$ \\
\hline
\multirow{4}{*}{NGC 5273}& \ion{He}{II} & $<0.35^{+2.17}$ &$<0.176^{+1.075}$ & \multirow{4}{*}{\cite{bentz2014mass}}\\
& \ion{H}{$\gamma$}  & $2.14^{+1.09}_{-1.08}$ &$0.456^{+0.237}_{-0.232}$ \\
& \ion{H}{$\beta$}  & $2.22^{+1.19}_{-1.61}$ &$0.443^{+0.245}_{-0.327}$ \\
& \ion{H}{$\alpha$}  & $2.06^{+1.42}_{-1.31}$ &$0.550^{+0.383}_{-0.353}$ \\
\hline
\multirow{4}{*}{3C 390.3}& \ion{He}{II} & $22.3^{+6.5}_{-3.8}$ &$11.4^{+4.0}_{-3.0}$ & \multirow{4}{*}{\cite{dietrich2012optical}} \\
& \ion{H}{$\gamma$}  & $58.1^{+4.3}_{-6.1}$ &$30.5^{+2.3}_{-3.3}$ \\
& \ion{H}{$\beta$}  & $44.3^{+3.0}_{-3.3}$ &$25.7^{+2.5}_{-2.7}$ \\
& \ion{H}{$\alpha$}  & $56.3^{+2.4}_{-6.6}$ &$25.7^{+2.0}_{-3.4}$ \\
\hline
\multirow{2}{*}{NGC 7469}& \ion{He}{II} & $1.3^{+0.9}_{-0.7}$ &$0.13^{+0.09}_{-0.07}$ & \multirow{2}{*}{\cite{peterson2014reverberation}} \\
& \ion{H}{$\beta$}  & $10.8^{+3.4}_{-1.3}$ &$0.34^{+0.13}_{-0.08}$ \\
\hline
\multirow{5}{*}{Mrk 142}
& \ion{He}{II}  & $<1.20^{+1.53}$ &$<1.2^{+1.9}$ & \multirow{5}{*}{\cite{bentz2010lick}}\\
& \ion{He}{I}  & $1.81^{+1.99}_{-1.20}$ &$0.24^{+0.28}_{-0.19}$ \\
& \ion{H}{$\gamma$}  & $2.86^{+1.22}_{-1.05}$ &$0.33^{+0.15}_{-0.13}$ \\
& \ion{H}{$\beta$}  & $2.74^{+0.73}_{-0.83}$ &$0.207^{+0.074}_{-0.080}$ \\
& \ion{H}{$\alpha$}  & $2.78^{+1.17}_{-0.88}$ &$0.248^{+0.109}_{-0.085}$ \\
\hline
\multirow{5}{*}{SBS 1116+583A}& \ion{He}{II} & $0.47^{+0.39}_{-0.47}$ &$0.27^{+0.22}_{-0.26}$ & \multirow{5}{*}{\cite{bentz2010lick}} \\
& \ion{He}{I}  & $<2.57^{+1.59}$ &$<0.88^{+0.56}$ \\
& \ion{H}{$\gamma$}  & $1.84^{+0.61}_{-0.51}$ &$0.37^{+0.13}_{-0.11}$ \\
& \ion{H}{$\beta$}  & $2.31^{+0.62}_{-0.49}$ &$0.55^{+0.20}_{-0.18}$ \\
& \ion{H}{$\alpha$}  & $4.01^{+1.37}_{-0.95}$ &$0.61^{+0.25}_{-0.18}$ \\
\hline
\multirow{5}{*}{Mrk 1310}& \ion{He}{II} & $0.94^{+0.61}_{-0.81}$ &$0.33^{+0.22}_{-0.29}$ & \multirow{5}{*}{\cite{bentz2010lick}} \\
& \ion{H}{$\gamma$}  & $1.82^{+0.63}_{-0.71}$ &$0.133^{+0.054}_{-0.059}$ \\
& \ion{He}{I}  & $2.56^{+0.90}_{-1.05}$ &$0.51^{+0.21}_{-0.24}$ \\
& \ion{H}{$\beta$}  & $3.66^{+0.59}_{-0.61}$ &$0.214^{+0.086}_{-0.086}$ \\
& \ion{H}{$\alpha$}  & $4.51^{+0.66}_{-0.61}$ &$0.238^{+0.061}_{-0.059}$ \\
\hline
\multirow{3}{*}{Mrk 202}& \ion{He}{II} & $1.47^{+2.40}_{-1.19}$ &$0.59^{+0.97}_{-0.49}$ & \multirow{3}{*}{\cite{bentz2010lick}}\\
& \ion{H}{$\gamma$}  & $3.31^{+1.57}_{-1.38}$ &$0.170^{+0.112}_{-0.095}$ \\
& \ion{H}{$\beta$}  & $3.05^{+1.73}_{-1.12}$ &$0.136^{+0.082}_{-0.057}$ \\
\hline
\multirow{4}{*}{NGC 4748}&\ion{He}{II} & $<1.0^{+7.8}$ &$<0.4^{+2.9}$ & \multirow{4}{*}{\cite{bentz2010lick}} \\
& \ion{H}{$\gamma$}  & $6.92^{+2.60}_{-3.22}$ &$0.58^{+0.29}_{-0.33}$ \\
& \ion{H}{$\beta$}  & $5.55^{+1.62}_{-2.22}$ &$0.25^{+0.10}_{-0.12}$ \\
& \ion{H}{$\alpha$}  & $7.50^{+2.97}_{-4.57}$ &$0.82^{+0.35}_{-0.52}$ \\
\hline
\multirow{5}{*}{NGC 6814}& \ion{He}{II} & $5.00^{+1.98}_{-1.83}$ &$3.4^{+1.4}_{-1.3}$ & \multirow{5}{*}{\cite{bentz2010lick}}\\
& \ion{He}{I}  & $3.09^{+1.33}_{-0.84}$ &$3.4^{+2.7}_{-2.5}$ \\
& \ion{H}{$\gamma$}  & $6.05^{+2.65}_{-2.34}$ &$0.98^{+0.47}_{-0.43}$ \\
& \ion{H}{$\beta$}  & $6.64^{+0.87}_{-0.90}$ &$1.76^{+0.33}_{-0.34}$ \\
& \ion{H}{$\alpha$}  & $9.46^{+1.90}_{-1.56}$ &$1.13^{+0.25}_{-0.22}$ \\
\hline

\hline
\hline
\end{tabular}
\end{table}
\endgroup

\section{Sample selection and Methodology}\label{sec:sample}
We obtain our sample \citep[using primarily The AGN Black Hole Mass Database from][]{bentz2015agn} by including every published RM study in which reliable black hole masses were obtained from emission lines of two or more ionization states. This results in our final sample of 14 objects reported in Table \ref{table:emdata}.
As most ground based RM studies are in the optical wavelength regime, the prominent lines covered in these analyses are: \ion{He}{II} 4686 \r{A}, \ion{He}{I} 5876 \r{A} as well as \ion{H}{$\alpha$}, \ion{H}{$\beta$} and \ion{H}{$\gamma$}. In different objects in our sample, these lines allow us to probe the regions deep within the gravitational well of the SMBH, between $\sim 8 \times $ $10^{-4}$ and $2.5 \times $ $10^{-2}$ pc. In some cases, the emission line measurements are accompanied with negative error bars that exceed the size of the central value (leading to unphysical values for $r_c$ and $M_{BH}$). In these cases, we treat the measurements as upper limits instead (based on their positive error), and do not include them in the statistical analysis. We show these upper limits using arrows in our figures. \par
For the mass distribution inferred from the multi-line determinations in our sample, we consider two models: (1) a single parameter model in which the enclosed mass (i.e., the effective black-hole mass) remains constant and (2) a two parameter model, which includes a dark matter spike as defined in equation~(\ref{eq:parametrization}) along with a constant black-hole mass. In the second model, the constant black-hole mass is fixed to be the mass obtained from the lag in the emission line closest to the black hole (except for the case of Mrk 1310, where the mass is obtained from the second closest line due to the large (but not unphysical) error bars for the first point that encompass all other data points). We note that different studies from which the $M_{BH}$ values have been compiled have used different values of the virial coefficient ($f$) in their analysis. As a result, the best-fit constant black hole mass in model 1 and the overall normalization in model 2 of each individual source will be subject to the same systematic uncertainty as $f$. However, all the studies use the same virial coefficient for each individual emission line in the same object. Therefore, the steepness ($\gamma$) of the DM component in model 2 is not affected by the choice of a particular value for $f$.
\par
We obtain the best-fit solutions to the mass distribution for the two models and compare them using the Bayesian Information Criterion \citep[BIC, see][]{1978AnSta...6..461S,Claeskens_Hjort_2008}. We account for the different degrees of freedom of the two models by defining the corrected $\chi^2$ as:

\begin{equation}
\chi^2_{\rm BIC}=\chi^2+k \, \mathrm{ln}(N)
\end{equation}
The difference between the $\chi^2_{\rm BIC}$ for the two models, defined as $\Delta \chi^2_{\text{BIC}}$ ($=\chi^2_{\text{BIC}_{\text{Spike}}} - \chi^2_{\text{BIC}_{\text{Const}}}$) is then a direct indicator of the preference of one model over the other. A negative $\Delta \chi^2_{\text{BIC}}$ favors a dark matter spike component, whereas a positive $\Delta \chi^2_{\text{BIC}}$ favors the constant black-hole mass scenario. In particular, we interpret (absolute) values greater than 9 as strong evidence for a given model, values between 2 and 9 are regarded as positive evidence, and values in between 0 and 2 as weak evidence \citep{Claeskens_Hjort_2008}.

\begin{figure*}[p]
\centering
\includegraphics[scale=0.60]{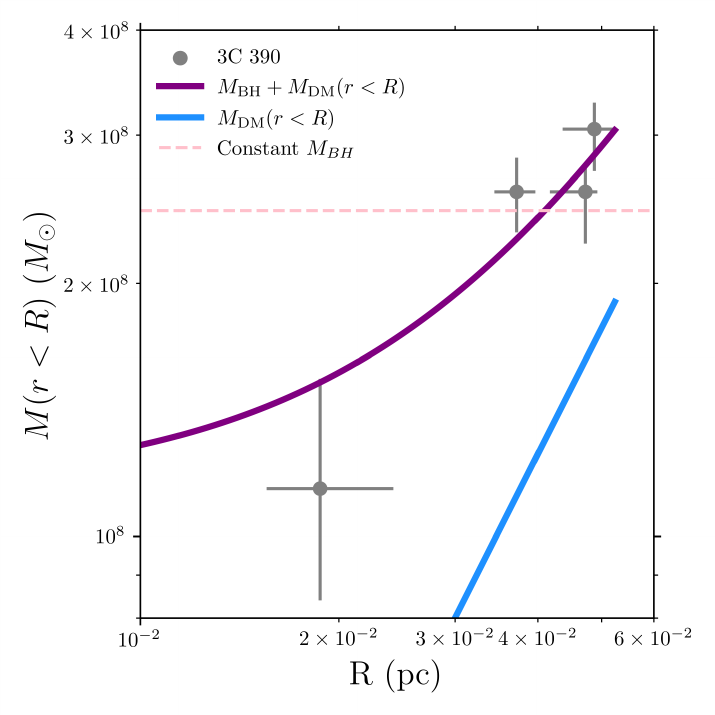}
\includegraphics[scale=0.60]{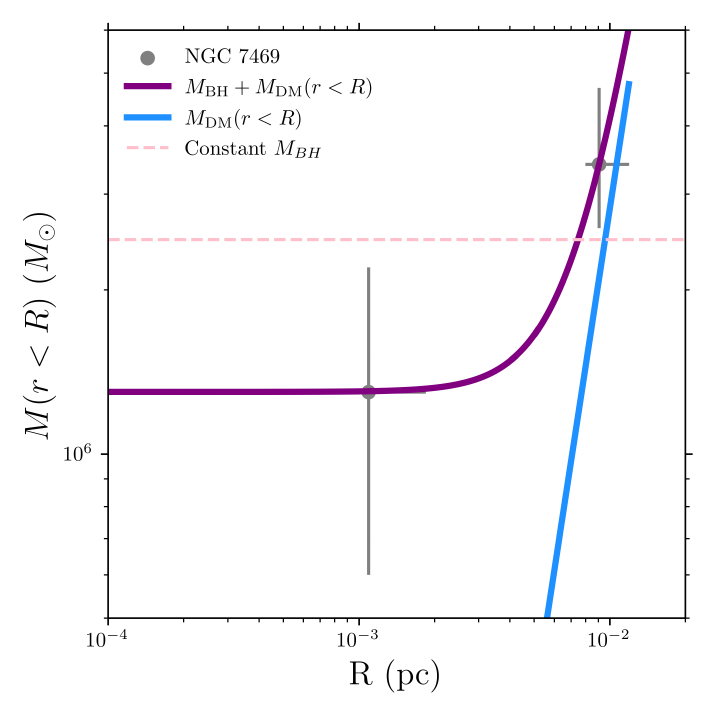}
\includegraphics[scale=0.60]{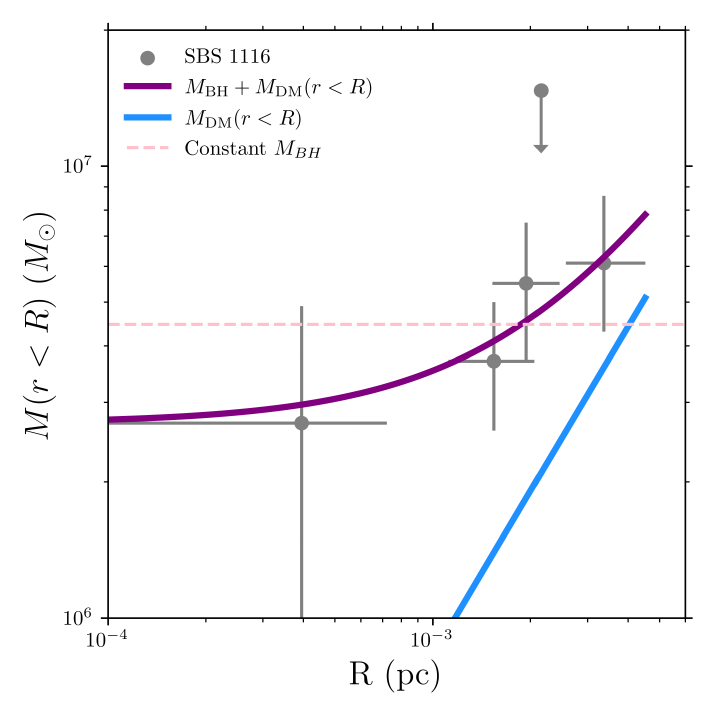}
\includegraphics[scale=0.60]
{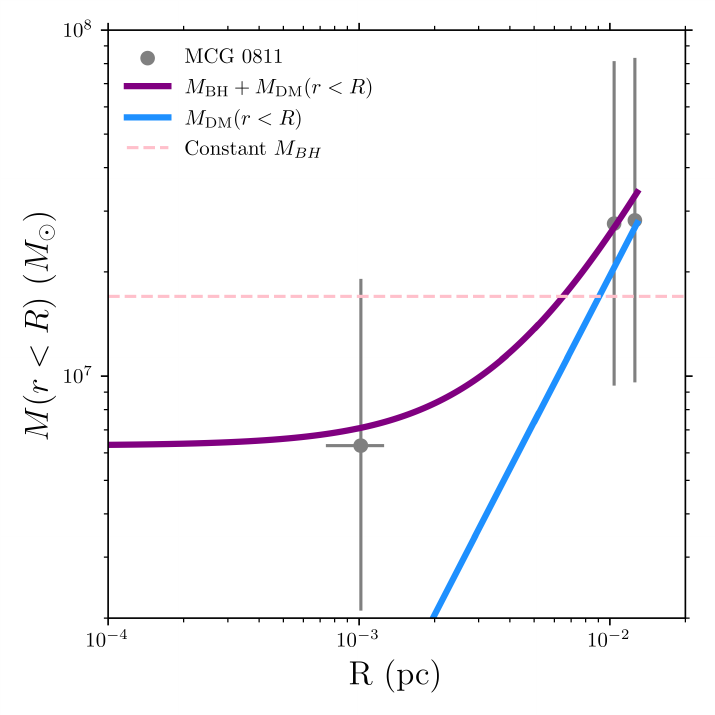}
\includegraphics[scale=0.60]{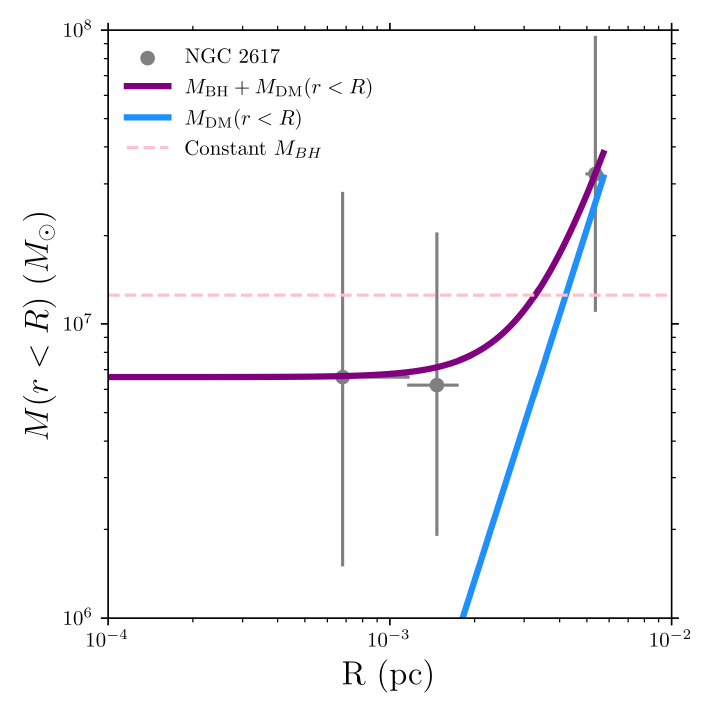}
\vspace{-0.20cm}
\caption{Measurements of the enclosed mass within different radii for five sources presenting differing levels of evidence for a dark matter component: strong (3C 390), positive (NGC 7469) and weak (SBS 1116, MCG 0811, NGC 2617) evidence (see section \ref{sec:results} and Fig. \ref{fig:confidencepositivestrong}). For comparison, we show the best fit obtained for a flat constant mass (pink dashed), and the best fit for an increasing mass component (purple solid), which includes contribution from the dark matter component (shown in blue).}
\label{fig:positivestrong}
\end{figure*}

\begin{figure*}[p]
\centering
\includegraphics[scale=0.6]{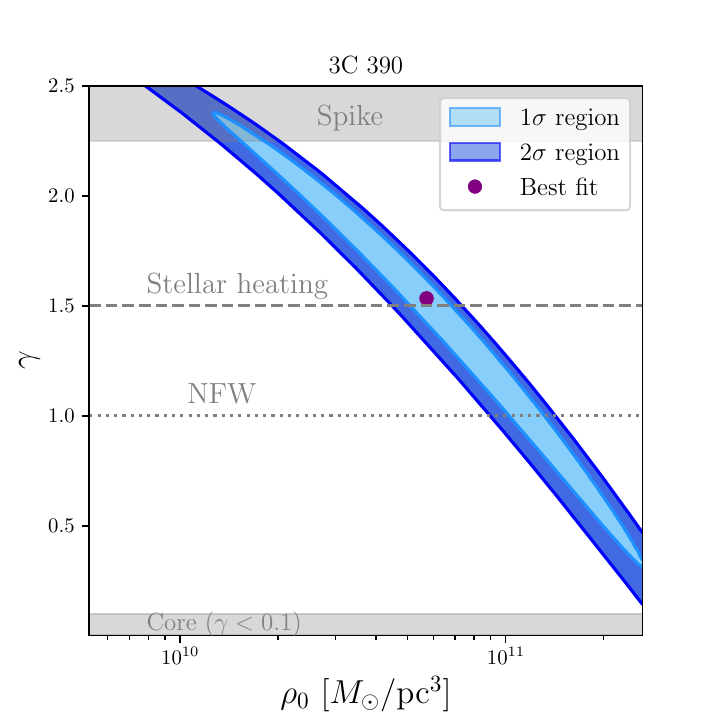}
\includegraphics[scale=0.6]{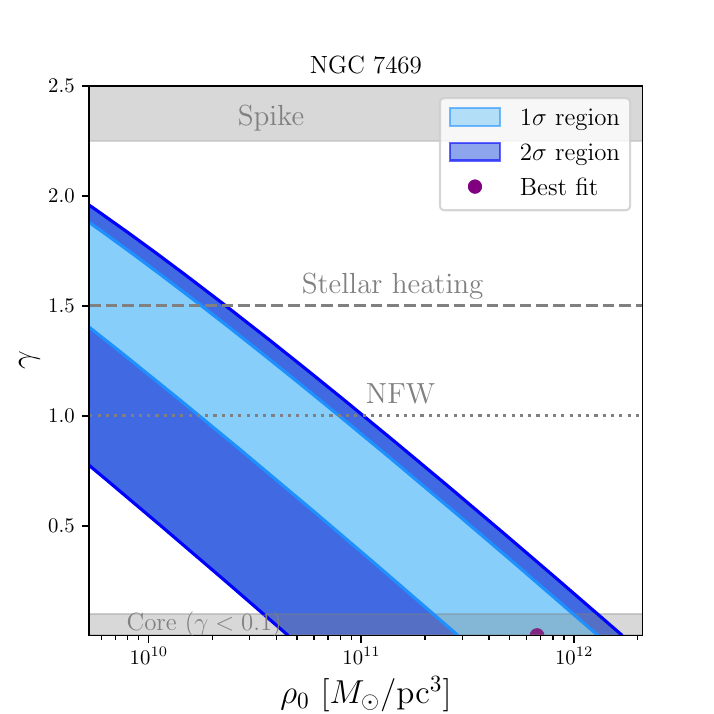}
\includegraphics[scale=0.6]{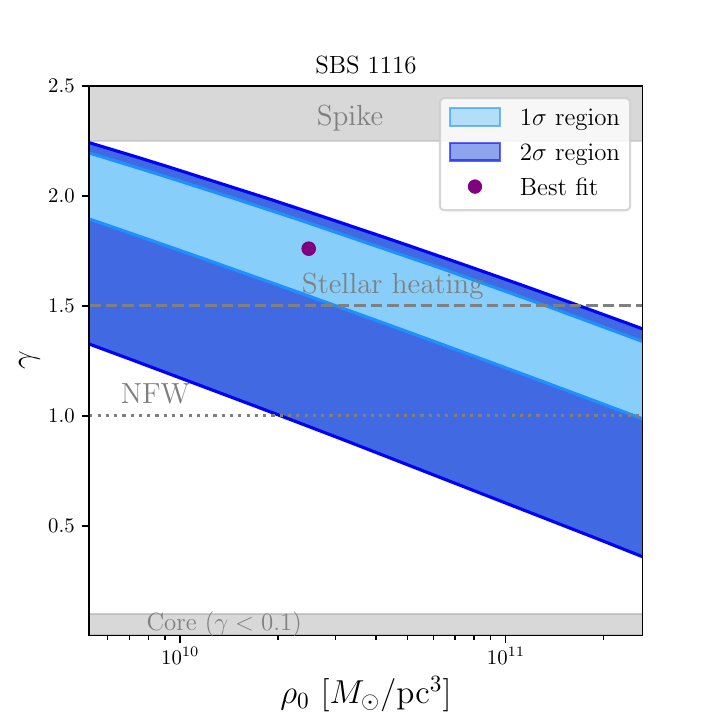}
\includegraphics[scale=0.6]{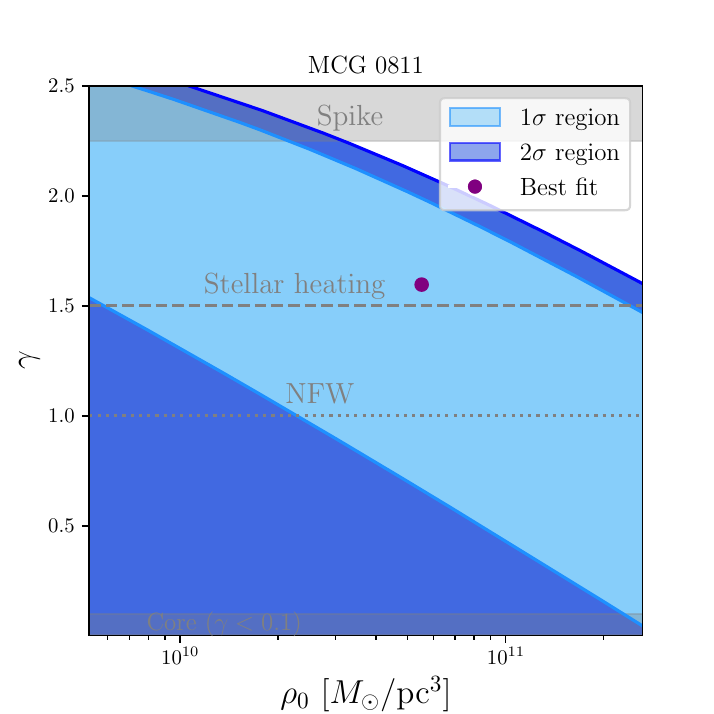}
\includegraphics[scale=0.6]{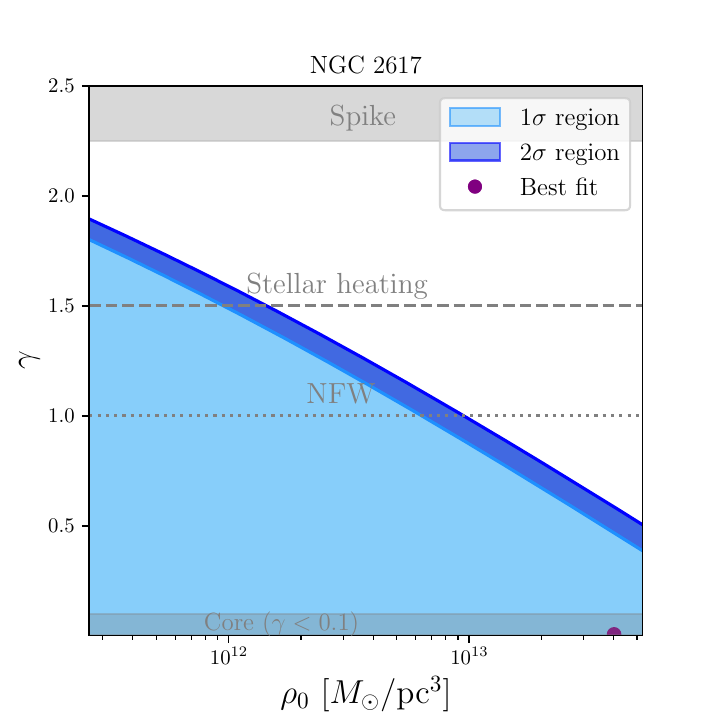}

\vspace{-0.2cm}
\caption{Confidence contours on the dark matter profile parameters---density $\rho_0$ and radial slope $\gamma$; see Eq.~(\ref{eq:parametrization})---for AGN with strong (3C 390), positive (NGC 7469) and weak (SBS 116, MCG 0811, NGC 2617) evidence. For comparison, we show in horizontal markings the expected profile indexes for different theoretical expectations: a dark matter spike, a spike relaxed by stellar heating, an NFW profile, and a dark matter core.}
\label{fig:confidencepositivestrong}
\end{figure*}

For sources with a preference for a dark matter component, we determine the 1$\sigma$ and 2$\sigma$ contours around the best fit solution in the ($\rho_0$, $\gamma$) parameter space. For this purpose, we perform a $\chi^2$ analysis, where the confidence contours are determined as
\begin{equation}
\chi^2(\rho_0, \gamma)-\chi^2_{\rm min} \leq \chi^2_{\sigma}
\end{equation}
where the $\chi^2$ distribution for two degrees of freedom is denoted as $\chi^2_{\sigma}$.

\section{Results}\label{sec:results}
Figure \ref{fig:positivestrong} shows the enclosed mass as a function of radii obtained from RM measurements (gray points), for five object (this figure is continued in the appendix for the remaining objects). The purple curve shows the mass profile obtained by adding a dark matter component (shown in blue) to the constant black hole mass inferred from the closest data point to the central black hole. The dashed-pink line shows the best-fit constant black hole mass value derived using all the RM measurements. We compare the two best-fit models for each object, with the results shown in Table \ref{tab:bic_comparison}. Out of the fourteen objects considered, two show strong/positive evidence for a DM spike (3C 390 and NGC 7469) and four show strong/positive preference for a constant black-hole mass (Mrk 110, 3C 120, Mrk 202 and NGC 6814). For the eight remaining objects, both models provide comparable fits (as $ |\Delta \chi^2_{\text{BIC}}| \lesssim$ 1), with a slight preference for a DM component in three and for a constant black-hole mass in five.  \par
For the five objects with a preference for a DM component ($ \Delta \chi^2_{\text{BIC}} <$ 0), we constrain the allowed parameter space for each source in Fig. \ref{fig:confidencepositivestrong}. The 1 and 2 $\sigma$ confidence regions around the best-fit solution (marked by the purple dot) are shown in light and dark blue respectively. For comparison we also mark the $\gamma$ values that are expected from dark matter spike formation scenarios, stellar heating processes, an NFW profile, and a profile that approaches a core. We notice that for some sources, the allowed regions can be quite narrow, favoring certain scenarios strongly over others. For instance, 3C 390 disfavors a dark matter cored profile and a very steep spike regardless of the normalization of the dark matter profile, and prefers a value of $\gamma \sim 1.5$, close to that expected from stellar heating processes. Similar, although weaker, conclusions can be drawn from SBS 1116. For MCG 0811, the preferred $\gamma$ is still close to the stellar heating predictions, but the allowed contours are too wide to meaningfully restrict the parameter space. On the contrary, NGC 7469 and NGC 2617 prefer a shallower index, closer to a core. In the Appendix, we show analogous plots for those objects with no evidence for a dark matter component (Fig. \ref{fig:noevidence_mass} and \ref{fig:noevidence_mass_2}), where we can still place an upper limit that disfavors the top-right corner of our parameter space (Fig. \ref{fig:Parameter_Limits}) with large values of $\gamma$ and $\rho_0$. 

We also report the best-fit fractional mass of each object in Table \ref{table:fractional}, together with the corresponding 1$\sigma$ errors. The errors on the fractional mass values were obtained by considering the parameter space enclosed by the 1$\sigma$ contours in Fig. \ref{fig:confidencepositivestrong}. While the parameter space does not appear to be closed at a 1$\sigma$ confidence level for any object except 3C 390, we are still able to obtain the errors by considering the entire possible range of $\rho_0$ corresponding to the theoretically motivated range of $\gamma=0-2.5$. Interestingly, the fractional mass of the DM component is found to be similar in all sources, agreeing with a value of $\sim 0.6$ within error bars, despite the wide range in the mass of the central SMBH for these objects.  

We further check whether there is a global preference for a particular range of $\gamma$ in the objects that show preference for a DM component. This is done by obtaining $\chi^2$ only as a function of $\gamma$ for these objects after marginalizing over $\rho_0$. We plot the behavior of the individual $\chi^2$ functions around their minima in Fig. \ref{fig:chi2_global}, where $\Delta\chi^2(\gamma) = \chi^2(\gamma) - \chi^2_{min}$. The combined $\Delta\chi^2$ function for all the sources is also shown in black and is found to have a well defined minima at $\gamma_{\rm global}\simeq1.6$, which is close to the expected value from a dark matter spike which has been depleted over time due to gravitational interactions with stars \citep{Merritt:2003qk, Gnedin:2003rj}. Qualitatively, we note that the major contribution to the observed behavior of combined $\Delta\chi^2$ comes from only two sources: 3C 390 and SBS 1116, particularly for smaller values of $\gamma$. For larger values of $\gamma$, the behavior is still dominated by 3C 390, with some minor contribution from all other sources. 

\begin{table*}[ht!]
\setlength{\tabcolsep}{4pt}
\renewcommand{\arraystretch}{1.10}
\centering
\caption{Best-fit parameters for the constant black hole mass model ($M_{BH}$) and the dark matter component ($\rho_0,\gamma$), along with their comparison.}
\label{tab:bic_comparison}
\begin{tabular}{lcccccccc}
\toprule
Source & Redshift & Morphology & $n$ & $M_{BH,\mathrm{best}}$ & $\rho_{0,\mathrm{best}}$ & $\gamma_{\mathrm{best}}$ & $\Delta \chi^2_{\mathrm{BIC}}$ & DM Evidence  \\
\midrule
3C 390.3  & 0.056 & S0& 4  & 24.40 & 5.5 $\times 10^{10}$ & 1.548 & -9.43 & Strong (in favor)\\
NGC 7469  & 0.016 & SBa& 2& 0.25 & 6.8 $\times 10^{11}$  & 0.001 & -2.35 & Positive (in favor) \\
SBS 1116  & 0.028 & SABa& 4  & 0.45 & 2.4 $\times 10^{10}$ & 1.764 & -0.37 & Weak (in favor)\\
MCG 0811  & 0.020 & SB0& 3 & 1.70 & 5.5 $\times 10^{10}$ & 1.594 & -0.29 & Weak (in favor)\\
NGC 2617  & 0.014 & SAa& 3 & 1.25 & 4.0 $\times 10^{13}$ & 0.001 & -0.03 & Weak (in favor)\\
\hline
Mrk 1310  & 0.020 & SA& 5 & 0.20 & 3.7 $\times 10^{10}$ & 1.313 & +0.35 & Weak (against)\\
NGC 4748  & 0.015 & Sa& 3 & 0.29 & 1.7 $\times 10^{12}$ & 0.001 & +0.55 & Weak (against)\\
Mrk 335  & 0.026 & S0a& 2 & 2.69 & 1.4 $\times 10^{11}$ & 0.031 & +0.68 & Weak (against)\\
NGC 5273  & 0.004 & SA& 3 & 0.47 & - & -  & +1.38 & Weak (against)\\
Mrk 142  & 0.045 & SBa& 4  & 0.24 & - & - & +1.38 & Weak (against)\\
Mrk 110  & 0.035 & Sa& 4  & 1.76 & - & - & +7.01 & Positive (against) \\
3C 120  & 0.033 & S0& 5 & 10.26 & - & - & +21.88 & Strong (against) \\
Mrk 202  & 0.023 & -& 3 & 0.16 & - & - & +44.95 & Strong (against)\\
NGC 6814  & 0.005 & SABbc& 5  & 1.35 & - & - & +129.27 & Strong (against) \\
\bottomrule
\end{tabular} \\
\justify \footnotesize \textbf{Note:} Redshift and morphology were taken from the NASA/IPAC Extragalactic Database. $n$ is the number of data points available from RM measurements. $M_{BH,\mathrm{best}}$ is in units of 10$^7 M_{\odot}$ and $\rho_{0,\mathrm{best}}$ is in units of $M_{\odot}$/pc$^3$. The last column reports the evidence inferred for/against a DM component and its strength based on the BIC criterion (see text for more details).
\label{table:BIC}
\smallskip
\end{table*}

\begingroup
\setlength{\tabcolsep}{5pt}
\renewcommand{\arraystretch}{1.50}
\begin{table}[ht]
\centering
\caption{Inferred dark matter component mass as a fraction of the total enclosed mass obtained with RM measurements at the farthest radii measured for each object.}
\vspace{0.3cm}
\begin{tabular}{lc}
\toprule
Source & $M_{\rm DM}(r<R_{\rm sp})/M_{\rm tot}(r< R_{\rm sp})$ \\
\midrule
3C 390.3 & $0.60_{-0.04}^{+0.04}$   \\
NGC 7469 & $0.62_{-0.21}^{+0.14}$  \\
SBS 1116 & $0.58_{-0.27}^{+0.15}$   \\
MCG 0811 & $0.81_{-0.57}^{+0.13}$\\
NGC 2617 & $<0.80^{+0.15}$ \\
\bottomrule
\end{tabular}
\smallskip
\label{table:fractional}
\end{table}

\begin{figure}[ht!]
         \centering
         \includegraphics[width=1.0\linewidth]{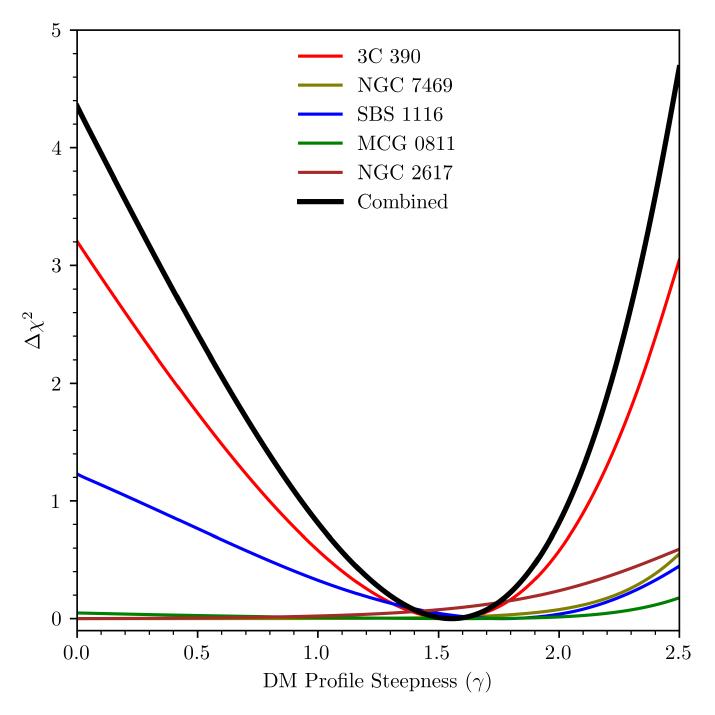}
         \caption{Individual $\Delta \chi^2$ as a function of $\gamma$ for the objects in our analysis presenting preference for a DM component. We also show the combined $\Delta \chi^2$ for these objects, which has a well defined minimum at $\gamma \simeq 1.6$.}
         \label{fig:chi2_global}
\end{figure}

\section{Discussion} \label{sec:discussion}

\subsection{Physical plausibility and implications of the derived DM spike parameters} \label{sec:physical}
The enclosed dark matter spike masses inferred from reverberation mapping data (see Table \ref{table:fractional}) are higher than expected in usual formation models \cite{Gondolo:1999ef}. It is worth then exploring, for these inferred spike parameters, what would be the consequences for the outer halo component characteristics. One restrictive boundary condition for the dark matter distribution in a galaxy is given by its halo mass. The total integrated dark matter in the spike and the outer halo profiles can not yield a mass exceeding the total halo mass of the galaxy. The halo mass for our sample can be estimated based on their SMBH mass, using the following scaling relation \cite{Ferrarese:2002ct, Baes:2003rt} 
\begin{equation}\label{eq:BH_Halo_Mass}
M_{\mathrm{BH}} \lesssim 10^8 M_{\odot}\left(\frac{M_{\mathrm{DM}}}{10^{12} M_{\odot}}\right)^{4 / 3}.
\end{equation}
We can then infer the outer halo parameters that when combined with the inner spike (described by the parameters in Table \ref{tab:bic_comparison}) will be consistent with the expected total halo mass of the host galaxies. We consider two different values for the outer spike radius of 0.01 and 0.1 pc \citep[following the expectations from semi-analytical models, e.g.,][]{Gondolo:1999ef, Ullio:2001fb, Merritt:2006mt, Merritt:2003qk} and find the corresponding parameters for the NFW-like profile that would satisfy the total halo mass obtained from equation \ref{eq:BH_Halo_Mass}. We perform this exercise for all sources in our sample that show a preference for a spike component, and show the resulting profiles in Fig. \ref{fig:inferred_total_profiles}. We find that the best fit parameters obtained with RM data for the inner distribution of dark matter around SMBHs would require very steep outer halo profiles in order to avoid overshooting the total mass of the halo. For the standard NFW profile
\begin{equation}
\rho(r)=\frac{\rho_0}{\frac{r}{r_s}\left(1+\frac{r}{r_s}\right)^2}
\end{equation}
with scale radius $r_s$, the edge of the halo is defined by the virial radius $R_{\rm virial}=cr_s$, where $c$ is the concentration parameter. For our sample, we find $r_s$ in the range $0.5-10$ pc, resulting in concentration parameter values between $c \simeq 10^{3}-10^{6}$.
These values are extreme compared to the expected values of $r_s \sim 20$ kpc and $c \sim 10$ for a Milky Way like galaxy. While active galaxies are indeed expected to have different halo parameters than local-quiescent galaxies \citep{booth2010dark, nino2025correlation}, this effect alone cannot explain the orders-of-magnitude difference seen here. This discrepancy is thus likely to be reflective of uncertainties in current RM based mass determinations, which we discuss in the next section.

\begin{figure*}[ht!]
\centering
\includegraphics[width=0.49\linewidth]{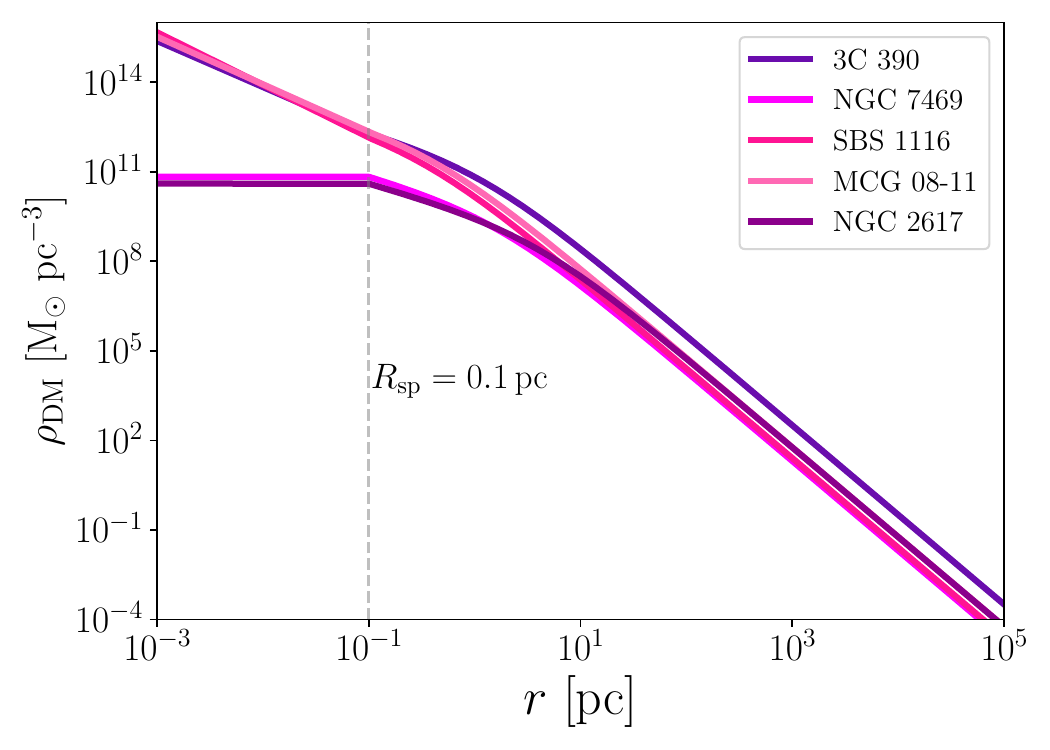}
\includegraphics[width=0.49\linewidth]{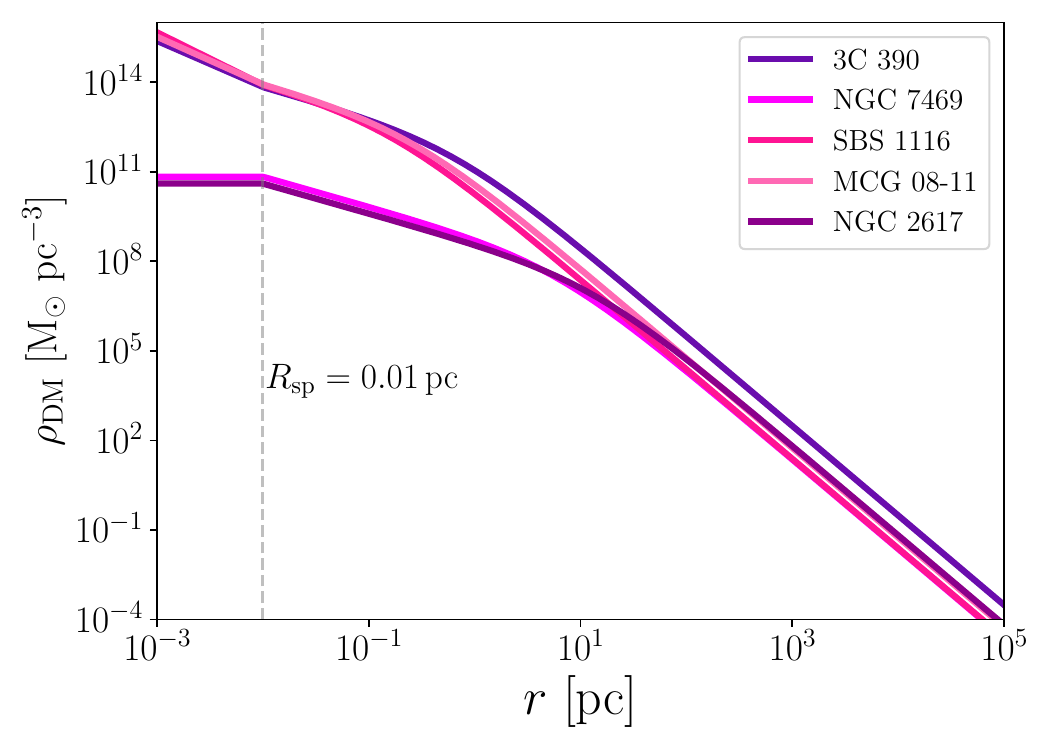}
\caption{Dark matter density profiles from sub-pc to kpc scales from the sources in our sample presenting evidence for an inner dark matter component. The inner density profile parameters are the best-fit values from Table \ref{table:BIC}, and outer halo parameters are fitted to reproduce the total mass of the galaxy expected from the phenomenological relation in equation \ref{eq:BH_Halo_Mass}.
The outer spike radius ($R_p$) is assumed to be 0.1 pc and 0.01 pc in the left and right panel, respectively.}
\label{fig:inferred_total_profiles}
\end{figure*}

\subsection{Uncertainty in RM based $M_{BH}$ determinations}

The mass of the black-hole determined from the lags observed in individual emission lines suffer from multiple layers of uncertainties. Statistical uncertainties depend primarily on the: (a) signal-to-noise of the emission lines, (b) cadence of the observations (i.e., temporal resolution) and (c) the overall duration of the campaign \citep[e.g.,][]{bentz2009lick}. These are included in the error bars reported in Table \ref{table:emdata} (and shown for all data points in Figs. \ref{fig:positivestrong}, \ref{fig:noevidence_mass}, \ref{fig:noevidence_mass_2}) and taken into account in our model comparison. \par
These measurements also suffer from large systematic uncertainties that are harder to quantify. Their effect is clearly seen in at least two objects in our sample (3C 120 and NGC 6814) where the measurements suggest a physically impossible decrease in enclosed mass, even when considering the large statistical error bars. These systematic uncertainties stem primarily from the unknown geometry and dynamics of the BLR in individual objects \citep[][]{krolik2001systematic}. Determining the enclosed mass from the observed lag measurements is dependent on the implicit assumption of a uniformly distributed and illuminated BLR, as well as an isotropic and virialized velocity distribution. Deviations from these highly idealized assumptions can significantly affect the derived mass estimates, even from different emission lines within the same object. Based on the results from Section \ref{sec:physical}, it is likely that these systematics affect other sources in our sample as well. The increasing mass profiles inferred from the individual RM measurements thus only represent an upper limit, as the difference in masses can also be ascribed to unknown systematics.\par
Fortunately, recent years have seen significant improvements in reducing the effects of these systematics. \cite{pancoast2011geometric} introduced a Bayesian framework which models the BLR as a collection of clouds, characterized by their position and velocity. In theory, this allows direct inference of the spatial and velocity distribution of the BLR from the data, reducing the uncertainties associated with the assumptions on its geometry and dynamics. Several other models have also been developed independently to improve upon other systematics as well by incorporating e.g., non-linear light responses and photoionization models \citep{li2013bayesian, rosborough2024modeling}. Constraining a growing number of parameters needed to accurately represent the actual physical conditions requires higher and higher data quality. These techniques have thus been applied so far to only a select few objects. In even fewer objects ($\approx$ 5), interferometry with the GRAVITY instrument at the Very Large Telescope has made it possible to spatially resolve the BLR and obtain a direct view of its size and dynamics and thus the enclosed mass \citep{gravity2018spatially, amorim2024size}. \par
Even with these recent advancements, however, the studies have only targeted a single emission line per object and worked under the implicit assumption of the enclosed mass at sub-parsec scales being only from the supermassive black hole. Without multi-line measurements that trace the mass profile over a range of distances, their impact on the study of dark matter distribution on sub-pc scales will remain limited. This, in turn, will affect the results of the RM and interferometry campaigns themselves, as the possible presence of an extended mass distribution on these scales, until/unless ruled out, will lead to a systematic overestimation of the inferred black-hole masses. 

\section{Conclusions}\label{sec:conclusions}
We have proposed a new method to constrain for the first time the dark matter density profile in the vicinity (at sub-parsec scales) of distant supermassive black holes, by making use of reverberation mapping measurements. For multiple AGN, RM measurements have been obtained for multiple emission lines with varying levels of ionization (e.g., He II, He I, H $\gamma$, H $\beta$ and H $\alpha$). Each of these lines can be used to infer, by means of the virial theorem, the enclosed mass within different distances from the central black hole, which allows us to map the radial mass density at these sources. 

For several sources in our sample, we find evidence for an increasing mass with radii, hinting at the presence of a dark matter component at sub-parsec distances. For those sources with a preference for a DM component over a constant black-hole mass model, we have derived confidence contours at the $1-2 \sigma$ level on the parameter space spanned by the normalization and steepness of the dark matter density profile at sub-parsec distances from the central black hole. We stress, however, that the majority of sources in our sample do not show preference for a model with increasing mass over a constant mass model.\par 
In the sample of sources that do have a preference for a DM component, current RM measurements point towards a preferred steepness of $\gamma \sim 1.6$, favored in scenarios where an initially formed dark matter spike is relaxed by stellar heating processes. They also suggest that in these objects the dark matter component can make up a large fraction of the total enclosed mass on the distance scales probed by RM, close to $\sim$ 60\%. This is larger than the expectations from standard theoretical formation scenarios. To match the total halo mass predicted from scaling relations, this also requires the outer halo profiles in these sources to be very steep, differing significantly from the profiles seen in local galaxies. \par
Our work establishes the first link between the observational technique of RM and the theoretical framework of dark matter spikes, both of which aim to study the same spatial scales in extragalactic SMBHs. While the current level of uncertainties in RM based mass measurements limit the conclusions that can be made about the dark matter distribution at sub-pc scales, significant advancements are being made in the field allowing for more accurate mass inferences. It is essential, however, that the scope of RM campaigns is widened, in general, to incorporate a larger number of emission lines in both observations and models. Parallel development
with dedicated simulations, going beyond current semi-analytical models into the distribution of dark matter at sub-parsec scales is also needed to confront the observational results with theoretical expectations.

\begin{acknowledgments}
We thank the anonymous referee for their constructive comments and suggestions that helped improve this paper.
M. S. and N. A. acknowledge support from NSF grant AST 2106249, as well as NASA STScI grants AR-15786, AR-16600, AR-16601 and AR-17556. The work of G. H. is supported by the U.S. Department of Energy under award
number DE-SC0020262. The work of S. H. is supported by NSF grant~PHY-2209420 and JSPS KAKENHI Grant Number JP22K03630 and JP23H04899. This work was supported by World Premier International Research Center Initiative (WPI Initiative), MEXT, Japan.
\end{acknowledgments}

\section*{Data Availability Statement}
The data underlying this article are compiled from publicly available literature. The specific references for each object are listed in Table~\ref{table:emdata}.

\bibliographystyle{unsrtnat}
\bibliography{apssamp}

\appendix

\section*{Appendix}
\begin{figure*}[t!]
\centering
\includegraphics[scale=0.6]{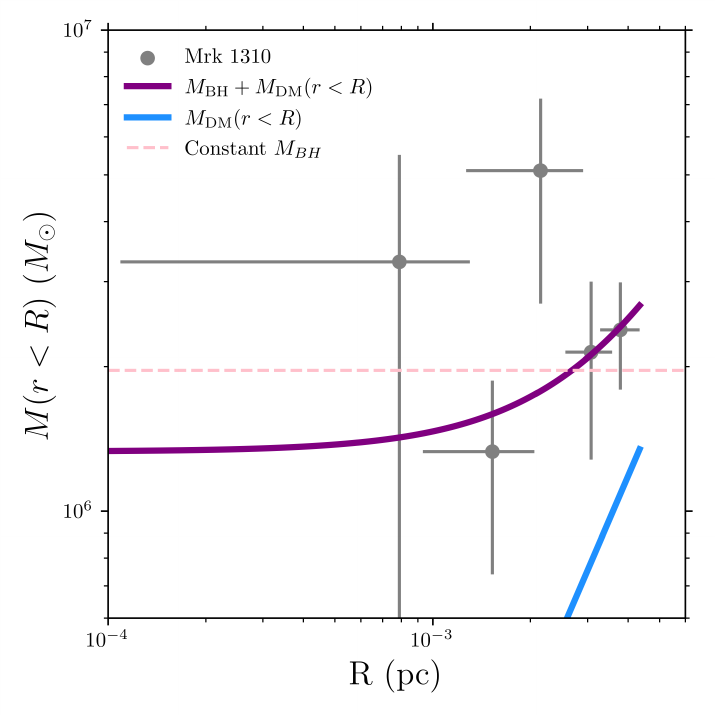}
\includegraphics[scale=0.6]{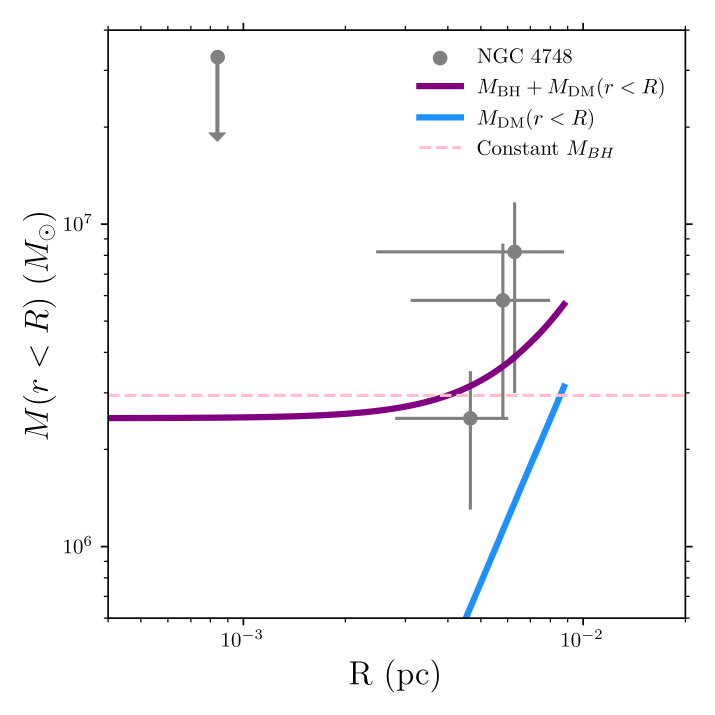}
\includegraphics[scale=0.6]{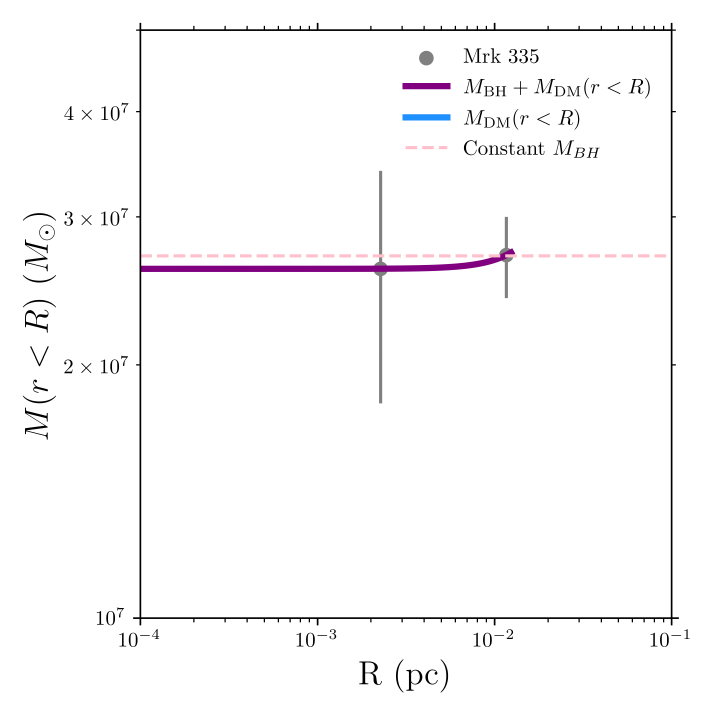}
\vspace{-0.2cm}
\caption{Measurements of the enclosed mass within different radii for sources showing an increasing mass, but no preference for dark matter component.}
\label{fig:noevidence_mass}
\end{figure*}

We report here the results of our model fitting procedure for the nine objects that did not show a preference for a DM component ($ \Delta \chi^2_{\text{BIC}} >$ 0). For three objects (Mrk 1310, NGC 4748 and Mrk 335), there is an increase in the enclosed mass with increasing radii (see Fig. \ref{fig:noevidence_mass}). Although our statistical BIC prescription (which accounts for the degrees of freedom of the model) does not allow us to claim any evidence for DM component in these objects, we note that it provides a better fit to the data than a constant mass scenario. We provide the best-fit parameters for both models for these sources in Table \ref{tab:bic_comparison}. For the remaining six objects, there is an apparent decrease in the enclosed mass beyond the closest point (see Fig. \ref{fig:noevidence_mass_2}). Thus the best fit when allowing for a positive density profile index still converges to a flat mass determined by the closest point. \par
We also derive upper limits on the dark matter profile parameters $(\rho_0, \gamma)$ from these sources. with the results shown in Fig \ref{fig:Parameter_Limits}. This was done by adopting the best-fit $M_{BH}$ obtained from model 1 as the input black-hole mass for model 2 and finding the parameter space excluded at a 1$\sigma$ confidence level based on the observed mass profile. We note that these upper limits disfavor scenarios with a steep dark matter profile index, as expected in canonical spike formation scenarios, but only for sufficiently large values of the normalization parameter $\rho_0$. In the range of parameters considered, the strongest constrain in this sample is obtained for Mrk 110.
\clearpage
\begin{figure*}
\centering
\includegraphics[scale=0.6]{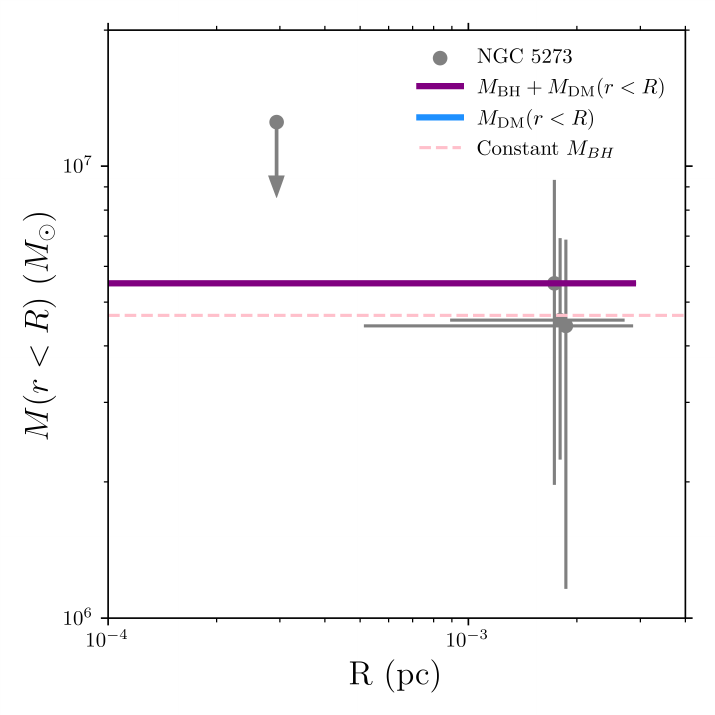}
\includegraphics[scale=0.6]{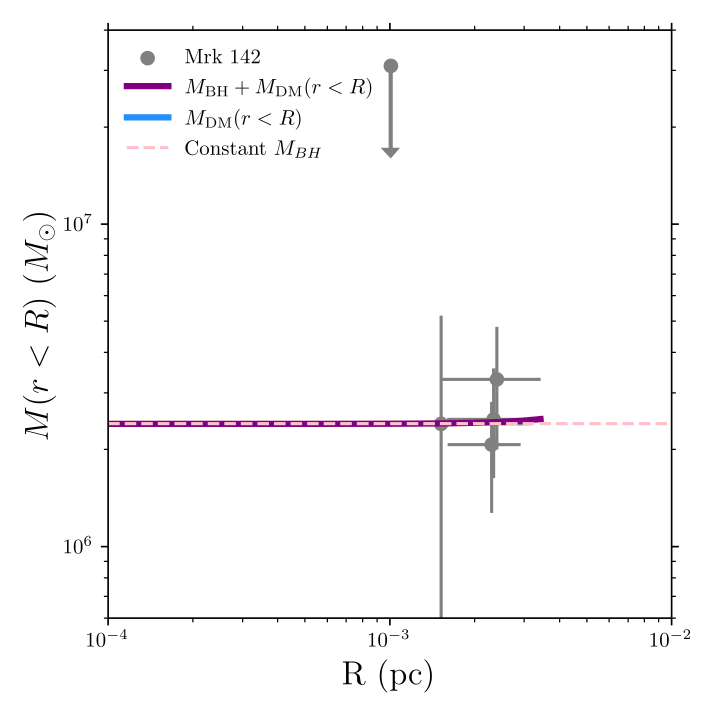}
\includegraphics[scale=0.6]{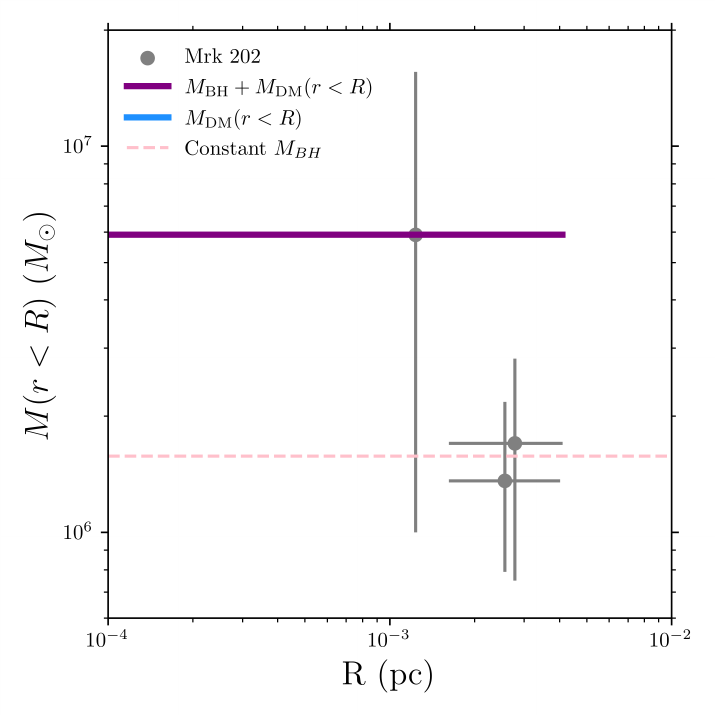} 
\includegraphics[scale=0.6]{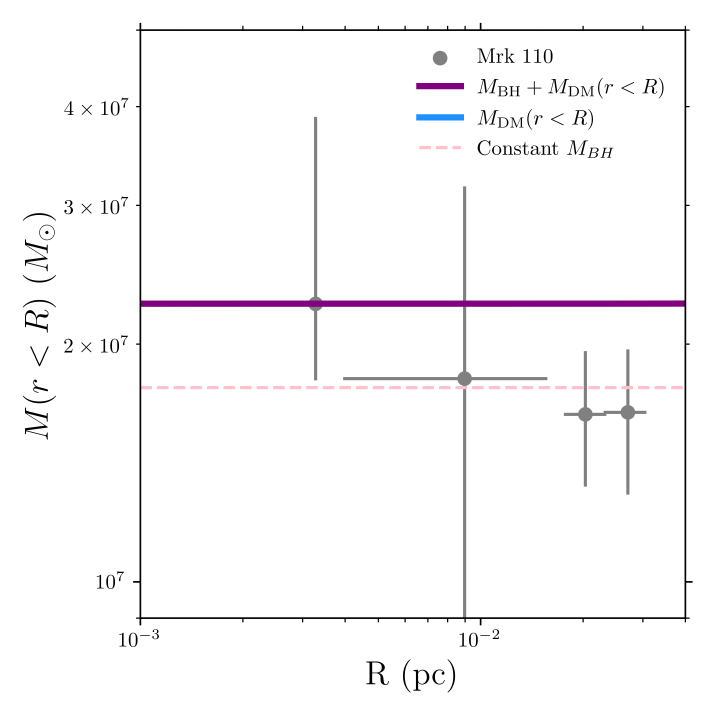}
\includegraphics[scale=0.6]{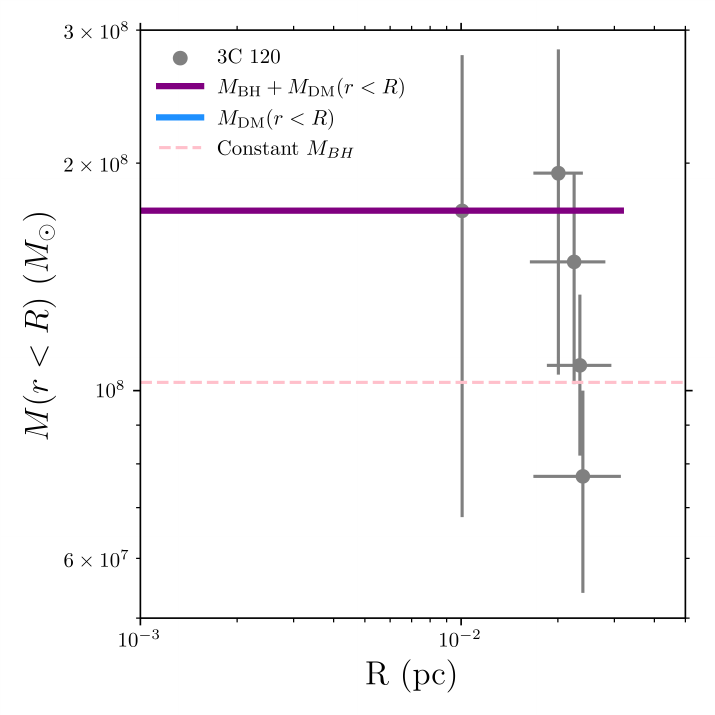}
\includegraphics[scale=0.6]{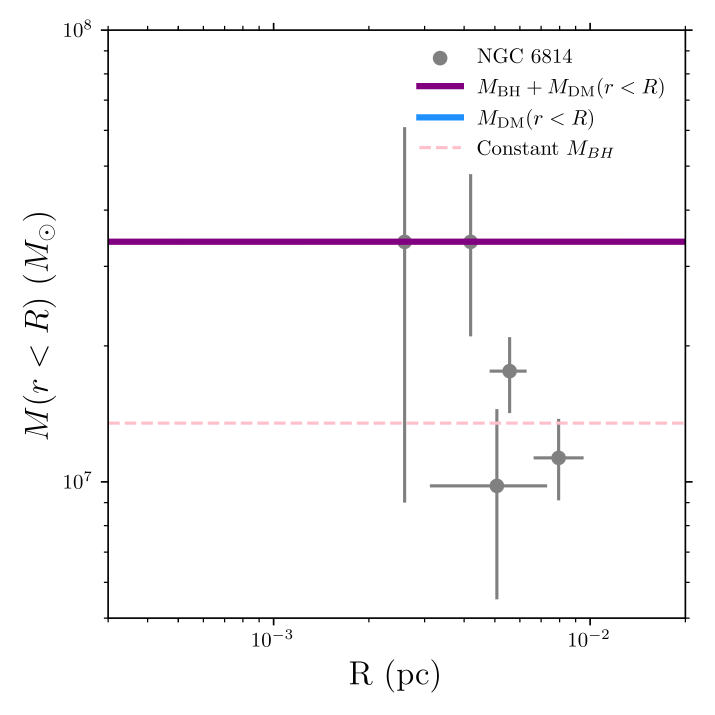}

\vspace{-0.2cm}
\caption{Measurements of the enclosed mass within different radii for sources presenting no evidence for a dark matter component.}
\label{fig:noevidence_mass_2}
\end{figure*}

\begin{figure}
\centering
\includegraphics[scale=0.8]{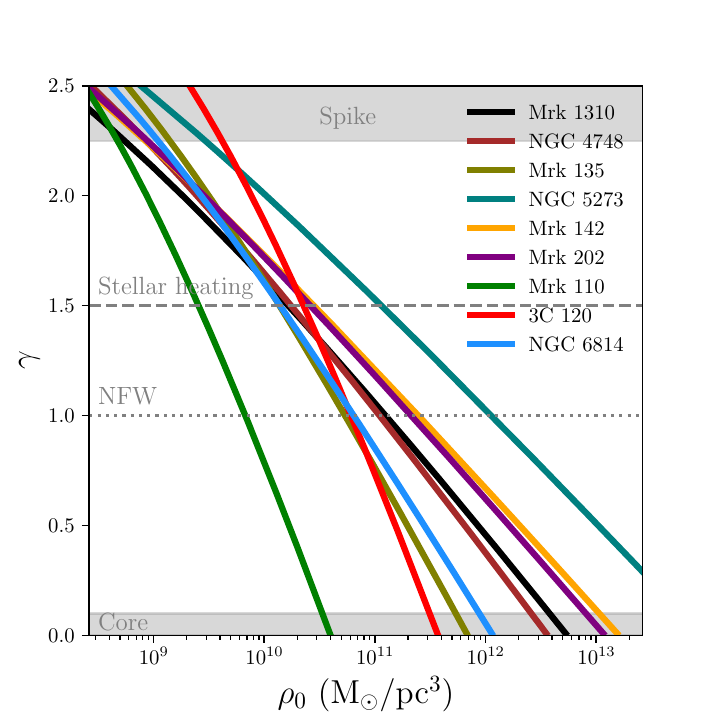}
\caption{Constraints on the $(\rho_0,\gamma)$ parameter space from objects with no evidence for a dark matter component under the BIC prescription. The parameter space above the individual curves is excluded with a 1$\sigma$ confidence.}
\label{fig:Parameter_Limits}
\end{figure}

\end{document}